\documentclass[
 aip,
 pop,
 amsmath,amssymb,
 superscriptaddress,
 reprint,%
  nofootinbib,
]{revtex4-1}

\usepackage{graphicx}
\usepackage{dcolumn}
\usepackage{bm}
\usepackage[dvipsnames]{xcolor}
\usepackage{amsmath}
\usepackage{enumerate}
\usepackage[draft]{hyperref}
\hypersetup{colorlinks=true}
\hypersetup{citecolor=blue}

\newcommand{\mvec}[1]{\mathbf{#1}}

\begin{document}

\preprint{AIP/123-QED}

\title[Nonlinear saturation of the Weibel instability]{Nonlinear saturation of the Weibel instability}

\author{P. Cagas}
  \affiliation{Virginia Tech, Blacksburg, VA 24060.}
\author{A. Hakim}
  \affiliation{Plasma Physics Laboratory, Princeton, NJ 08544.}
\author{W. Scales}
  \affiliation{Virginia Tech, Blacksburg, VA 24060.}
\author{B. Srinivasan}
  \email{srinbhu@vt.edu}
  \affiliation{Virginia Tech, Blacksburg, VA 24060.}
\date{\today}

\begin{abstract}
  The growth and saturation of magnetic fields due to the Weibel
  instability (WI) have important implications for laboratory and
  astrophysical plasmas, and this has drawn significant interest
  recently.  Since the WI can generate a large magnetic field from no
  initial field, the maximum magnitudes achieved can have significant
  consequences for a number of applications.  Hence, an understanding
  of the detailed dynamics driving the nonlinear saturation of the WI
  is important.  This work considers the nonlinear saturation of the
  WI when counter-streaming populations of initially unmagnetized
  electrons are perturbed by a magnetic field oriented perpendicular
  to the direction of streaming.  Previous works have found magnetic
  trapping to be important\cite{Davidson1972} and connected electron
  skin depth spatial scales to the nonlinear saturation of the
  WI.\cite{Califano1998} Results presented in this work are consistent
  with these findings for a high-temperature case. However, using a
  high-order continuum kinetic simulation tool, this work demonstrates
  that, when the electron populations are colder, a significant
  electrostatic potential develops that works with the magnetic field
  to create potential wells.  The electrostatic field develops due to
  transverse flows induced by the WI, and in some cases is
  strengthened by a secondary instability.  This field plays a key
  role in saturation of the WI for colder populations. The role of the
  electrostatic potential in Weibel instability saturation has not
  been studied in detail previously.
\end{abstract}

\maketitle

\section{Introduction}

The Weibel instability (WI)\cite{Fried1959, Weibel1959} has been
studied as a leading mechanism for the origin and growth of magnetic
fields in a number of laboratory\cite{Fox2013, Okada2007, Silva2002,
  Califano1997} and astrophysical plasma\cite{Lazar2009, Ghizzo2017}
applications.  Note that especially in the regime when drift
velocities are larger than thermal velocities, this instability is
also referred to as the current filamentation instability (CFI).  WI
can generate a large magnetic field from no initial field and can
amplify a small existing field by many orders of magnitude.  Hence,
the WI has generated a significant amount of interest in the
laboratory and astrophysics communities in recent years and a
comprehensive study of the growth and nonlinear saturation of the WI
is critical to estimate the saturated magnetic field magnitudes that
may be achieved.  Previous works have emphasized the role of magnetic
trapping in the nonlinear saturation of the WI\cite{Davidson1972} and
relate saturation to when the effective electron gyroradius in the
generated magnetic field becomes of the order of the electron
collision-less skin depth.\cite{Califano1998} In the work presented
here, using fully kinetic simulations, it is shown that for relatively
cold beams, in addition to the magnetic potential, an electrostatic
potential develops and plays a critical role in saturating the
WI. Note that in this context, the magnetic potential does not refer
to the vector potential, \textbf{A}, but rather to the integral of the
magnetic part of the Lorentz force, $\int
\left(\mathbf{v}\times\mathbf{B}\right)_xdx$.

The WI is studied in this work using two counter-streaming populations
of electrons perturbed by a magnetic field perpendicular to the beam
longitudinal direction of both beams.  The thermal velocity may be
varied with respect to the drift velocity to understand how the
saturation of the instability changes across these regimes.  Results
from two regimes are presented here.  The first is when the
counter-streaming velocity is smaller than the thermal velocity, the
system is analogous to a single population with anisotropic
temperature; the second is for the case when the counter-streaming
velocity is larger than the thermal velocity of two distinct
counter-streaming populations.  The magnetic component of the Lorentz
force (referred to as the ``filamentation force'' in this work) causes
the two populations to repel each other resulting in exponential
growth of the magnetic field.  For hot populations, the saturation of
the magnetic field occurs due to magnetic trapping, consistent with
previous work \cite{Davidson1972}.  The work presented here shows that
the saturation of the magnetic field for cold populations occurs due
to the formation of potential wells that counter the filamentation
force and halt the growth of the WI.  The wells are caused by a
combination of the magnetic and electrostatic potentials.

\section{Problem Description} \label{sec:setup}

To understand the nonlinear physics of the WI, continuum kinetic
simulations of the WI in one configuration space dimension and two
velocity dimensions (1X2V) are performed. The continuum kinetic model
uses the discontinuous Galerkin (DG)\cite{Cockburn2001} scheme with
serendipity basis set\cite{Arnold2011} to directly discretize the
Vlasov-Maxwell equations.  The Vlasov equation is solved for the
electron species,
\begin{align}\label{eq:vlasov}
  \frac{\partial f}{\partial t} + \textbf{v}\cdot\frac{\partial
    f}{\partial\textbf{x}} + \frac{q}{m}\left(\textbf{E} +
  \textbf{v}\times\textbf{B}\right)\frac{\partial
    f}{\partial\textbf{v}} = 0,
\end{align}
where $f$ is the distribution function, $q/m$ is the charge-to-mass
ratio, $\mathbf{E}$ and $\mathbf{B}$ are the electric and magnetic
fields evolved using Maxwell's equations.  Ions are considered
stationary in the time scales of the interest and are used only as a
non-evolving neutralizing background.  Extensive benchmarks are
presented in a companion numerics paper\cite{Juno2017}.  The base
method conserves energy exactly, meaning
\begin{multline}
  \frac{\partial}{\partial t}\left(\frac{1}{2}\iint mv^2 f(\mathbf{x},
  \mathbf{v}, t) d\mathbf{v}d\mathbf{x}
  \right.+\\+\left. \frac{1}{2}\int\epsilon_0 E^2 d\mathbf{x} +
  \frac{1}{2}\int\frac{B^2}{\mu_0} d\mathbf{x}\right) = 0.
\end{multline}
Tests performed to evaluate the energy conservation properties of this
algorithm\cite{Juno2017} show that the relative energy change is on
the order of $10^{-11}$.  However, the limiter, which is included to
ensure positivity of the distribution function, leads to small (of
order one percent) energy conservation errors. 

The WI simulations are initialized using two electron streams of
uniform density and temperature  along the $x$-axis
and uniform drift along the $y$-axis. 
This initial uniform, but unstable, equilibrium is disturbed with a
perturbation in $B_z$ given by,
\begin{equation}
  B_z(x) = B_{z,0}\mathrm{sin}(k_0 x),
\end{equation}
where the $k_0$ is the initial perturbation wave-number and
$B_{z,0}=10^{-3}$ in the dimensionless units. Note that
$B_{z,max}/B_{z,0} \approx 200$.  The configuration space ($x$-axis)
is periodic and ranges from 0 to $2\pi/k_0$, therefore, the initial
perturbation is exactly one sine wave.  In this work, results for
$k_0\lambda_D = 0.04$ are presented.  Simulations using higher $k$
have been performed to verify that the results shown here are
consistent for the range of unstable mode numbers. (For sufficiently
high mode numbers the WI is stable.)

The results presented here summarize findings that rely on the subtle
interplay between the magnetic and electric fields that leads to
nonlinear saturation of the WI. Hence, the ability to obtain a smooth,
noise-free phase-space distribution function is critical.

\section{Linear theory}

In order to obtain the kinetic dispersion relation, the Vlasov
equation (\ref{eq:vlasov}) is linearized
\begin{multline}
  -\mathrm{i}\omega f_{1_s} + \mathrm{i} v_xk_xf_{1_s} + \\\frac{q_s}{m_s}\left[(E_x+v_yB_z)\partial_{v_x}
    f_{0_s} + (E_y-v_xB_z)\partial_{v_y} f_{0_s}\right]=0,
\end{multline}
where $f_0$ is the equilibrium distribution function and $f_1$ is the
distribution function perturbation.

$f_0$ is the Maxwellian distribution function.  The perturbation is
combined with the linearized Ampere's law,
\begin{equation}
-\mathrm{i}k_xB_z =
\mu_0q \left[\int_\mathcal{V} v_yf_{1_1}\thinspace
  \mathrm{d}^2\mvec{v} + \int_\mathcal{V} v_yf_{1_2}\thinspace
  \mathrm{d}^2\mvec{v}\right]- \frac{\mathrm{i}\omega}{c^2} E_y.
\end{equation}

Elimination of the fields from the previous equation then leads to
the following kinetic dispersion relation,
\begin{equation}\label{eq:weibel:dispersion}
 \frac{1}{2} = \frac{\omega_0^2}{c^2k^2} \left[\zeta
  Z(\zeta)\left(1+\frac{u_d^2}{v_{th}^2}\right)
  +\frac{u_d^2}{v_{th}^2}\right]+\frac{v_{th}^2}{c^2}\zeta^2,
\end{equation}
where $\omega_0$ is the plasma oscillation frequency, $c$ is the speed
of light, $k$ is the instability wave-number, $u_d$ is the drift speed
of each population (assuming symmetric drift velocities $u_d$ with
respect to zero), and $v_{th}$ is thermal speed.  $\zeta =
\omega/(\sqrt{2}v_{th}k)$, where $\omega = \omega_r +
\mathrm{i}\gamma$. $Z(\zeta)$ is the plasma dispersion function
defined as
\begin{equation}
  Z(\zeta)
  = \frac{1}{\sqrt{\pi}} \int_{-\infty}^{\infty}\frac{e^{-x^2}}{x-\zeta}
   \mathrm{d}x.
\end{equation}
In the cold fluid limit, using the asymptotic expansion of $Z(\zeta)$
for large $\zeta$, the cold fluid Weibel dispersion relation is obtained
\begin{align}
  \frac{{\omega}^{4}}{2{k}^{2}} - \left( \frac{1}{2} + \frac{1}{{k}^{2}} \right)
  {\omega}^{2} -u_d^{2}  = 0, \label{eq:weibel:cold-dispersion}
\end{align}
where $\omega$ is normalized to $\omega_0$, $k$ is normalized to
$\omega_0/c$ and $u_d$ is normalized to $c$. Eq.\thinspace 12 in
Ref\thinspace[\onlinecite{Califano1997}] obtains the same cold fluid
dispersion relation described by
Eq.\thinspace\ref{eq:weibel:cold-dispersion} for the case of two
counter-streaming, but otherwise identical electron beams.  The cold
fluid dispersion relation predicts a \emph{larger} growth rate
compared to the growth rate that is obtained from the kinetic
dispersion relation in Eq.\thinspace\ref{eq:weibel:dispersion}.

\begin{figure}[!htb]
  \includegraphics[width=\linewidth]{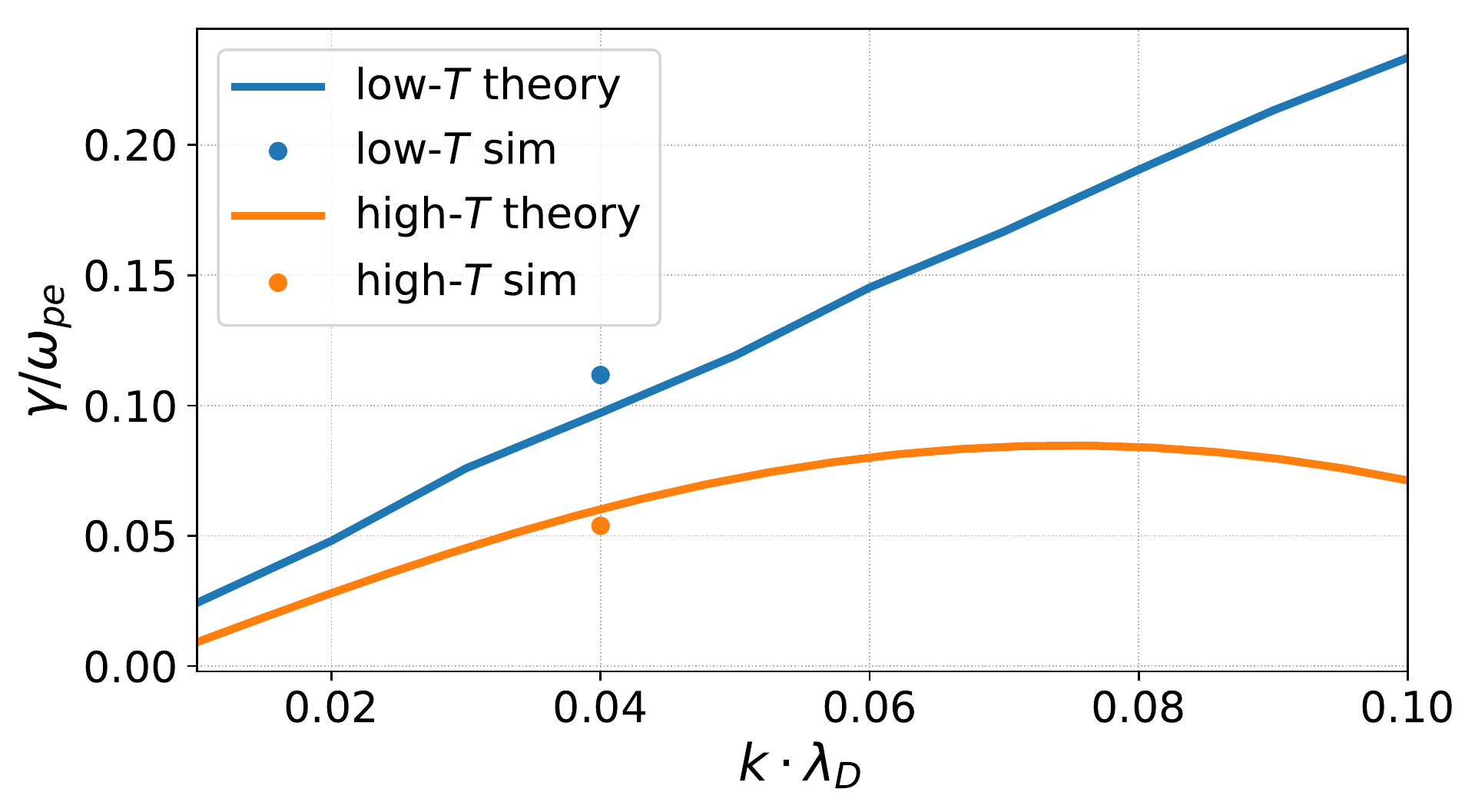}
  \caption{Linear theory prediction
    (Eq.\thinspace\ref{eq:weibel:dispersion}) of the Weibel
    instability growth rates for the two cases, a high temperature
    case ($u_d/v_{th}=1$) and a low-temperature case
    ($u_d/v_{th}=0.1$) as described in Table\thinspace\ref{tab:sim}.
    The solid line represents the linear theory results.  Dots
    represent the growth rates obtained from simulation.}
  \label{fig:weibel}
\end{figure}

Previous work\cite{Cagas2017} presents some initial results of the
linear growth of the WI benchmarked to kinetic theory using the same
continuum kinetic framework used here. The linear growth rates
obtained from the kinetic simulations are in reasonably good agreement
with theory for lower wavenumbers but begin to differ from theory for
higher wavenumbers.  The difference between theory and simulations for
high wavenumbers is likely due to plasma heating which occurs as the
instability grows and phase-space mixing occurs.  This heating is not
accounted for in the linear dispersion relation where the temperature
is assumed constant.  This is in agreement with the results discussed
in this work in Fig.\thinspace\ref{fig:phasespace}. Numerical values
of the growth rates for the two cases used in this work together with
the simulation results are presented in Fig.\thinspace\ref{fig:weibel}.

The growth rates from simulation are calculated by fitting the
simulation data to an exponential function (with 2 free parameters).
This is done by gradually increasing the region (in time) that is used
to calculate the fit.  The fit with the highest coefficient of
determination, $R^2$, is then selected and used to determine the
simulation growth rate.

\section{Continuum kinetic simulation results}\label{sec:results}

Simulation results are presented for two cases, a high-temperature
conuter-streaming population of electrons and a low temperature
version.  Using the problem description outlined in
Sec.\thinspace\ref{sec:setup}, the first case uses $v_{th} = u_d$,
which is relevant to the classical Weibel instability configuration,
while the second features distinct electron streams with $v_{th} <
u_d$.

Overview of the simulations is in the Table\thinspace\ref{tab:sim}.
Note that the speed of light is set to be artificially lower to
overcome time-step limitations.  Relativistic effects are not
considered in this work.
\begin{table}
  \begin{tabular}{cccc}
    \hline
    & $u_d/c$ & $v_{th}/c$ & $k_0\lambda_D$ \\
    High-temperature beams & $\pm$0.3 & 0.3 & 0.04 \\
    Low-temperature beams & $\pm$0.3 & 0.031 & 0.04 \\
    \hline
  \end{tabular}
  \caption{Overview of the simulations}
  \label{tab:sim}
\end{table}

\subsection{High-temperature beams}

\begin{figure}[!htb]
  \centering \includegraphics[width=\linewidth]{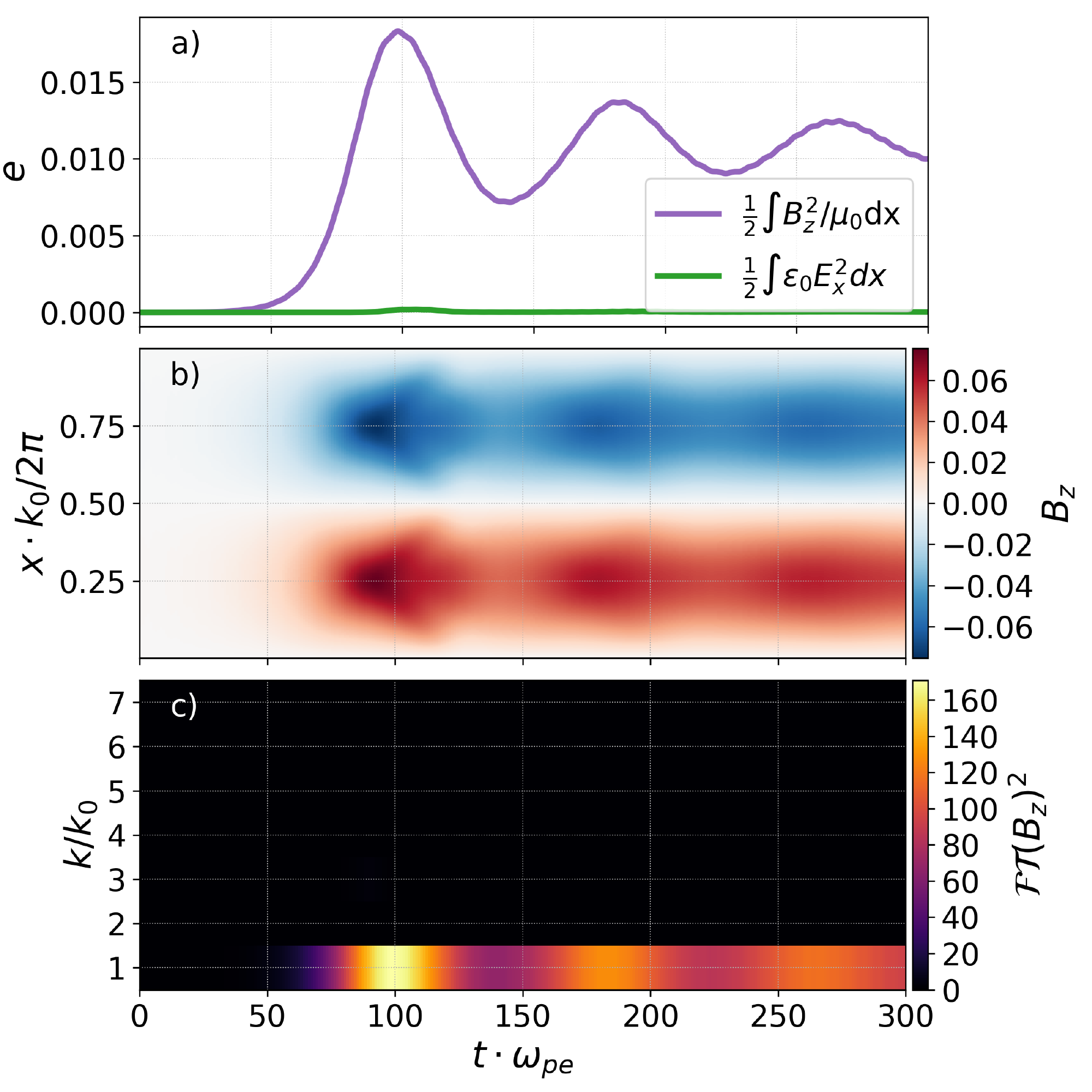}
  \caption{Evolution of the magnetic field in the high-temperature
    beams case.  Panel a) shows the growth of both total magnetic
    energy $\int B_z^2/(2\mu_0)dx$ with the strong periodic behavior
    after nonlinear saturation and the electric field energy $\int
    \varepsilon_0E_x^2/2dx$; panel b) shows the time evolution of
    $B_z$ for each $x$; finally, panel c) shows the spectrogram of
    $B_z$.  It is clear from the panel c) that only the mode seeded
    through the initial perturbation is growing.}
  \label{fig:highT_evolution}
\end{figure}

Simulations are performed for the high-temperature case using the
parameters described in Table\thinspace\ref{tab:sim} for the hot
electron beam.  A magnetic field grows significantly from the initial
perturbation.  Figure\thinspace\ref{fig:highT_evolution}(a) presents
the magnetic field energy and the electric field energy as a function
of time.  Note the exponential growth of the magnetic field energy
followed by nonlinear saturation.  There is negligible growth of the
electric field energy in comparison to the magnetic field energy for
this case.  Also note the distinct periodic behavior of the magnetic
field energy in the nonlinear phase of the instability.  The period of
the oscillations in the nonlinear part of the instability is
\begin{equation*}
  \frac{\omega}{\omega_{pe}} \approx \frac{2\pi}{82} = 0.076,
\end{equation*}
which compares very well to the theoretical magnetic bounce period
\cite{Davidson1972}
\begin{equation}
  \frac{\omega_B}{\omega_{pe}} = \sqrt{k\frac{q}{m}u_yB_z} \approx
  0.077,
\end{equation}
when the values of $k=0.4$, $u_y=0.3$, and $B_z=0.5$ are used.  These
values are consistent with the normalized parameters presented in
Table\thinspace\ref{tab:sim} for the hot electron case.  The
high-temperature case provides excellent agreement with previous
results \cite{Davidson1972} showing that magnetic trapping is the
primary mechanism for saturation of this instability.
Figures\thinspace\ref{fig:highT_evolution}b) and c) present the
magnetic field and its Fourier transform as a function of the
1-dimensional space (in the y-axis) and time (in the x-axis).  Note
that the magnetic field spatio-temporal profile shows that only a
single-mode grows throughout, and this is the same as the initialized
mode.  No other modes are growing in this case.

\subsection{Low-temperature beams}

\begin{figure}[!htb]
  \centering
  \includegraphics[width=\linewidth]{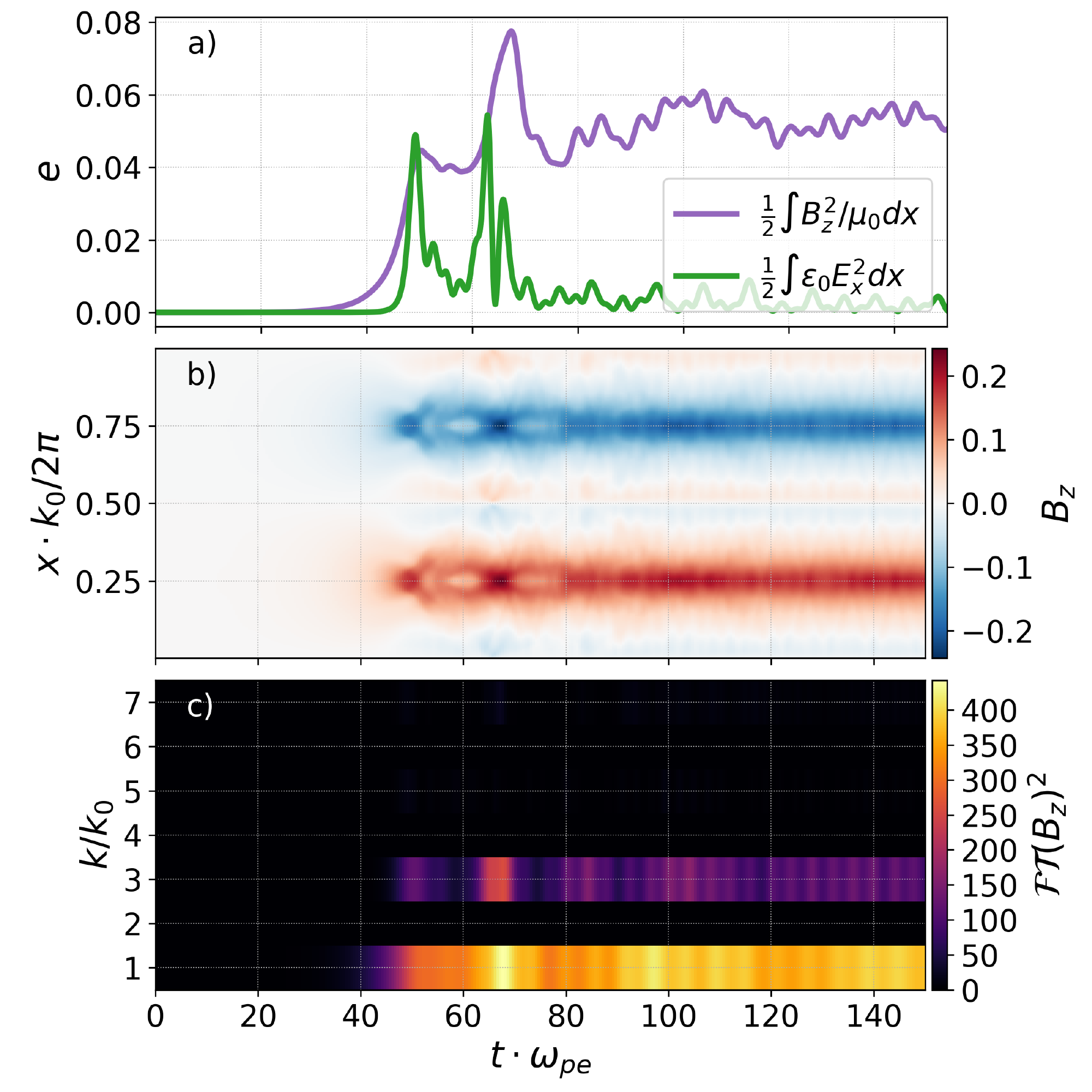}
  \caption{Evolution of the magnetic field in the colder beam case.
    Panel a) shows the growth of both total magnetic energy $\int
    B_z^2/(2\mu_0)dx$ with the strong periodic behavior in the
    nonlinear part and the electric field energy $\int
    \varepsilon_0E_x^2/2dx$; panel b) shows the time evolution of
    $B_z$ for each $x$; finally, panel c) shows the spectrogram of
    $B_z$.}
  \label{fig:lowT_evolution}
\end{figure}

Simulations are performed for the low-temperature case using the
parameters described in Table\thinspace\ref{tab:sim} for the cold
electron beam.  The evolution of the magnetic field energy and
electric field energy are presented in
Fig.\thinspace\ref{fig:lowT_evolution}(a).  There are key differences
in these results compared to the high-temperature case.  Firstly,
there is significant growth of the electric field energy, which is of
the order of the magnetic field energy at saturation.  Secondly, the
magnetic field energy contains a similar low frequency periodic
behavior in the nonlinear phase, but there is also a higher frequency
perturbation superimposed on top of it (note that the continuum kinetic
simulations used do not produce statistical noise like
particle-in-cell codes do).
Figures\thinspace\ref{fig:lowT_evolution}b) and c) present the
spatio-temporal profile of the magnetic field and its Fourier
transform similar to the high-temperature case.  An additional key
difference from the high-temperature case observed from the
spectrogram includes the presence of higher-k modes (compared to the
single initialized mode) that occur approximately around the time of
nonlinear saturation.

The filamentation force is introduced in the $x$-direction owing to
the drift $u_y$ (note that this work distinguishes bulk velocity $u$
and local velocity $v$; the bulk velocities for populations with
positive $u_y$ and negative $u_y$ are denoted as $u_y^+$ and $u_y^-$)
and the magnetic field perturbation $B_z$.  This force, $qv_y^\pm
B_z$, results in a transverse flow, $u_x$, in opposite directions for
each of the electron populations.  This filamentation leads to an
exponential growth of the magnetic field and the corresponding
magnetic field energy as seen in
Figs.\thinspace\ref{fig:highT_evolution} and \ref{fig:lowT_evolution},
which show magnetic and electric field energies.  Both energies are
converted from the free kinetic energy of the electrons.  It is worth
noting that due to the dimensions (1X2V) and problem setup, only the
$B_z$ component of the magnetic field grows.  For electric fields,
both $E_x$ and $E_y$ grow, but the energy corresponding to the $E_y$
component is over an order of magnitude lower than the energy
corresponding to $E_x$.

\begin{figure}[!htb]
  \centering
  \includegraphics[width=\linewidth]{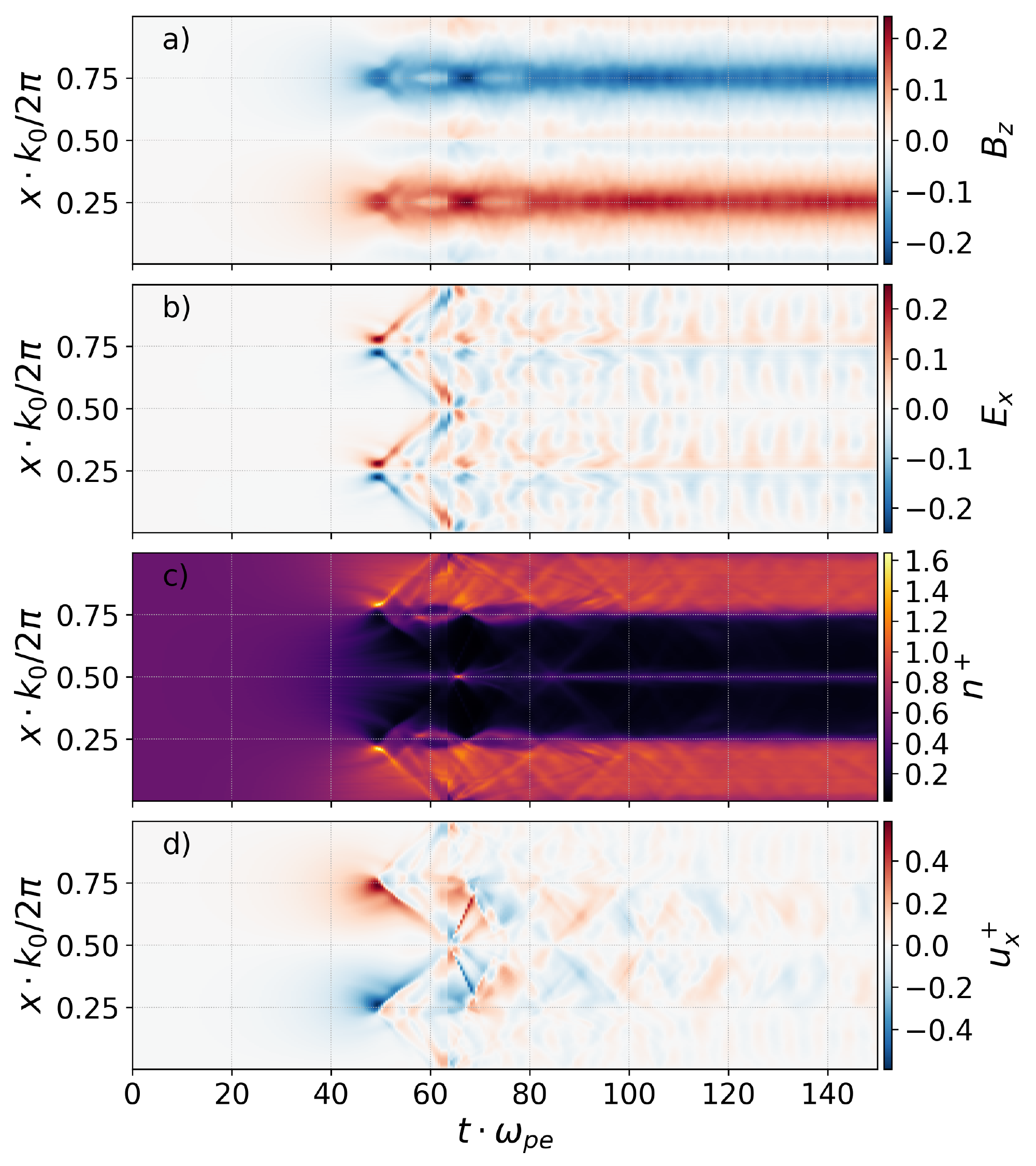}
  \caption{Evolution of the magnetic field $B_z$ (a), electric field
    $E_x$ (b), and number density (c) and $x$-velocity of the
    population with $u_y>0$.  Lower temperature case simulation with
    $v_{th}/u_d = 0.1$.}
  \label{fig:lowT_1}
\end{figure}

The evolution of $E_x$, $B_z$, density, and the $u_x$ velocity is
presented in Fig.\thinspace\ref{fig:lowT_1} as a function of space and
time.  More precisely, the bulk velocities and densities for each of
the populations with positive $u_y$ and negative $u_y$ are computed
separately.  Figure\thinspace\ref{fig:lowT_1} presents the results for
the population with $u_y > 0$, and spatial symmetry of the populations
can be used to understand the profile of the population with $u_y <
0$.  During the time of linear growth, a $u_x$ velocity develops
(Fig.\thinspace\ref{fig:lowT_1}d) and density flows from one part of
the domain to the other (Fig.\thinspace\ref{fig:lowT_1}c).  However,
unlike the high-temperature case, there is a significant increase in
the electric field (Fig.\thinspace\ref{fig:lowT_1}b) (mechanisms for
this will be discussed later in more detail) at the time of
instability saturation which rapidly stops the flow and consequently
saturates the growth of the instability.  Without the $x$-flow, the
electric field decays, and the $u_x$ flow is reintroduced by the
filamentation force.  The second saturation occurs soon after.  Figure
\ref{fig:lowT_1}d also provides information about the nonlinear
periodic phase.  When $u_x^+ > 0$ (red) in the region of $0.5 < x\cdot
k_0/2\pi < 1.0$ and negative in the other half, the particles flow
from the low-density to the high-density region increasing the
filamentation. Note that the second population behaves in the opposite
manner, i.e. $n^-$ has a maximum where $n^+$ has a minimum and vice
versa.  Because of this, the currents of the counter-streaming
populations do not cancel out and $B_z$ is growing.  On the other
hand, when $u_x^+ < 0$ (blue) in the region of $0.5 < x\cdot k_0/2\pi
< 1.0$ and positive in the other half, the gradients of density are
decreasing, currents cancel out, and magnetic field decreases as well.
Comparison of Fig.\thinspace\ref{fig:lowT_evolution}a and
Fig.\thinspace\ref{fig:lowT_1}d shows that this direction of $u_x$
corresponds well to regions where the magnetic field is increasing and
decreasing. Particles are moving from the lower density region to the
higher density region until the first saturation ($t\cdot\omega_{pe}
\approx 50$) and the flow is stopped.  At $t\cdot\omega_{pe}
> 60$, the flow is reintroduced and magnetic field is rising
again.  After the second saturation, the flow is reversed and magnetic
field is decreasing until $t\cdot\omega_{pe} \approx 75$, which is the
local minimum of the magnetic field energy, and the process repeats.
This behaviour is also visible in Fig.\thinspace\ref{fig:tsgrowth}.

\begin{figure}[!htb]
  \centering
  \includegraphics[width=\linewidth]{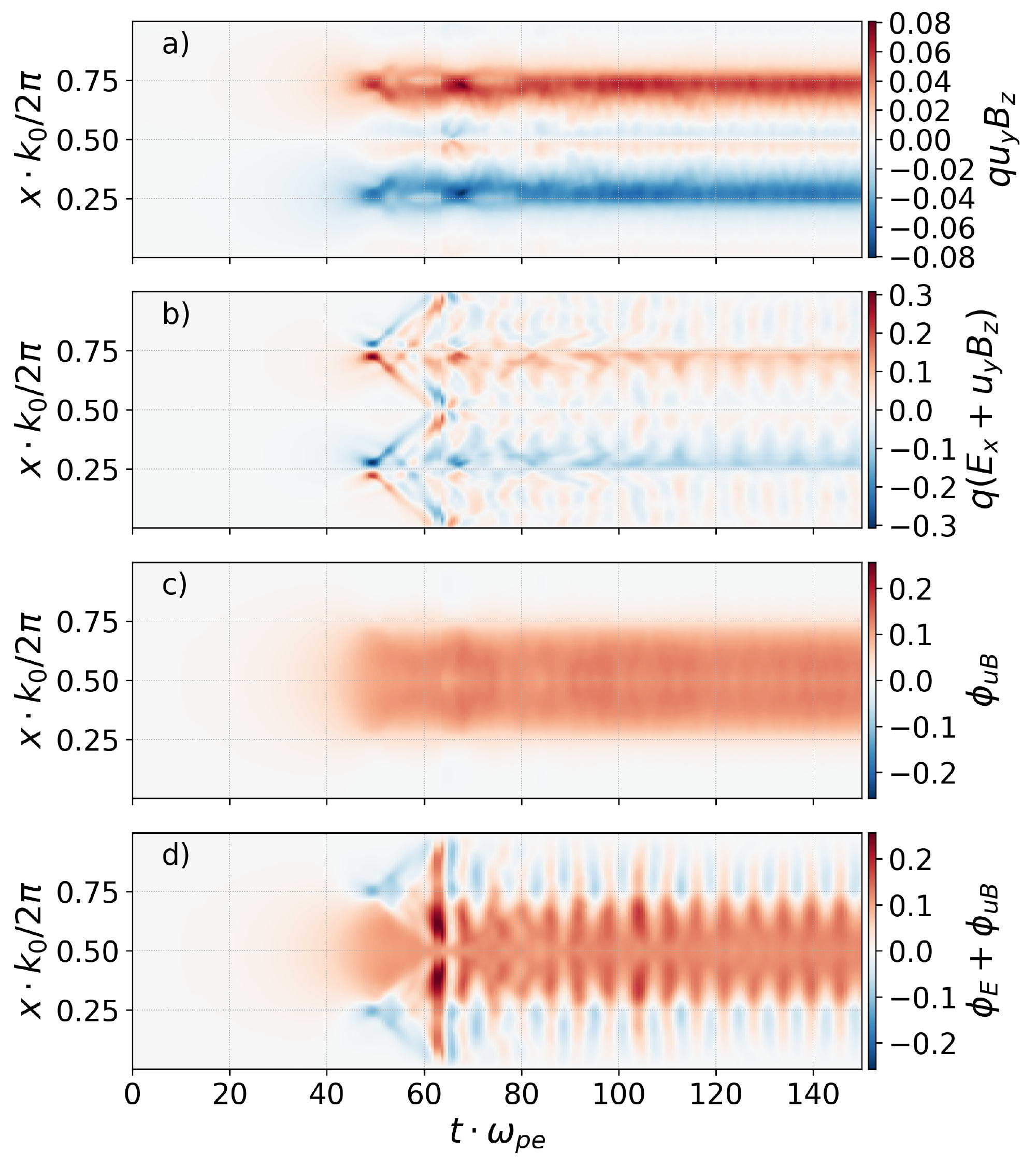}
  \caption{Evolution of the magnetic part of the Lorentz force
    (filamentation force) (a), full Lorentz force (b), and potentials
    corresponding to both of them (c, d).  The potentials here are
    calculated as the integrals of the forces.  Bulk velocities of the
    population with $u_d>0$ are used for this calculation.  Lower
    temperature case simulation with $v_{th}/u_d = 0.1$.}
  \label{fig:lowT_2}
\end{figure}

Figure\thinspace\ref{fig:lowT_2} provides additional insight into the
saturation through electromagnetic forces and the corresponding
potentials.  Note that, as was mentioned previously, the potential
does not refer to the magnetic vector potential, $\textbf{A}$, but
rather to the integral of the forces.  In other words,
\begin{equation}
  q \mathbf{u}\times\mathbf{B} = -\nabla \phi_{uB}.
\end{equation}

Figure\thinspace\ref{fig:lowT_2}a presents only the magnetic part,
$\mathbf{u}\times \mathbf{B}$, of the Lorentz force, whereas
Fig.\thinspace\ref{fig:lowT_2}b shows the full Lorentz force including
the non-negligible electric field contribution as well.
Figures\thinspace\ref{fig:lowT_2}c and d present the potentials
(integrals of the forces) corresponding to
Figs.\thinspace\ref{fig:lowT_2}a and b, respectively.  As expected,
the filamentation force potential, $\phi_{uB}$, creates a potential
well over one half of the domain ($x\cdot k_0/2\pi < 0.25$ and $x\cdot
k_0/2\pi > 0.75$; note that the domain is periodic in $x$), which is
consistent with the magnetically trapped particles bouncing between
magnetic field extremes (see Fig.\thinspace\ref{fig:lowT_2}c).
However, for this cold-temperature case, the electric field is
significant in the Lorentz force.  This is seen in
Fig.\thinspace\ref{fig:lowT_2}d) which describes modifications to the
overall potential well located at the boundary between the two
populations (maxima of the electric field).  The wells in these
boundary regions are narrower in comparison to the potential due to
the filamentation force alone, but with comparable depth (as noted
from the magnitudes).  If magnetic trapping was the sole mechanism in
this instability, the potential described by
Fig.\thinspace\ref{fig:lowT_2}c would represent the trapping
potential.  However, note the presence of a $\phi_E$ potential in
Fig.\thinspace\ref{fig:lowT_2}d due to the $E_x$ that develops and
grows.  The net result of these two potentials shows the net regions
of particle trapping as a function of time.  Hence, the electric field
trapping plays a significant role along with the magnetic trapping.

The electric field periodically rises and decays (see either
Fig.\thinspace\ref{fig:lowT_1}b or \ref{fig:lowT_2}d) which is
consistent with the high frequency oscillations in
Fig.\thinspace\ref{fig:lowT_evolution}a.

\subsection{Origin of the electric field}

The transverse flows introduced by the magnetic field coupled with the
nonuniform densities of the populations ($n^+(x) \neq n^-(x)$) are the
main source of the electric field.  Since $B_y=B_x=0$ and $\partial
B_z/\partial y = 0$, Ampere's law in the $x$-direction reduces to
\begin{equation}
  \varepsilon_0 \frac{\partial E_x}{\partial t} = -j_x =
  e\left[n^+u_x^+ + n^-u_x^-\right].
\end{equation}

The growth of the electric field can be estimated with
\begin{widetext}\begin{align}
  E_x(t) &= \frac{e}{\varepsilon_0} \int \left[n^+u_x^+ + n^-u_x^-\right] dt \\
  &= -\frac{e}{\varepsilon_0} \int \left[n^+
    \int\frac{e}{m}\left(E_x+u_y^+B_z\right)dt +
    n^-\int\frac{e}{m}\left(E_x+u_y^-B_z\right)dt\right] dt \\
  &\approx -\frac{e^2u_{y0}^+}{\varepsilon_0m} \int\left[
  \left(n^+-n^-\right) \int B_zdt \right]dt -
  \frac{e^2}{\varepsilon_0m}\int\left[\left(n^++n^-\right)\int E_x
    dt\right]dt \\
  &\approx -\frac{e^2u_{y0}^+}{\varepsilon_0m\gamma} \int\left[
  \left(n^+-n^-\right) C e^{\gamma t}\right]dt -
  \frac{e^2}{\varepsilon_0m}\int\left[\left(n^++n^-\right)\int E_x
    dt\right]dt \label{eq:egrowthcontd}
\end{align}\end{widetext}
where $\gamma$ is the magnetic field growth rate.  Using the
continuity equation and that $u_x \propto e^{\gamma t}$, one can
estimate the maximum $n^+ - n^-$ to be of order $e^{\gamma t}$.  The
contribution of the second term on the right-hand-side of
Eq. \ref{eq:egrowthcontd} is bounded and cannot be an exponential.
Using these approximations, an upper limit of the electric field is
obtained
\begin{align}
  |E_x(t)| &< C_1 e^{2 \gamma t} + C_2 \label{eq:egrowth}
\end{align}
and $C_1$ and $C_2$ are constants of integration.  This shows that the
electric field should be constrained by twice the exponential growth
of the magnetic field.

\begin{figure}[!htb]
  \centering
  \includegraphics[width=\linewidth]{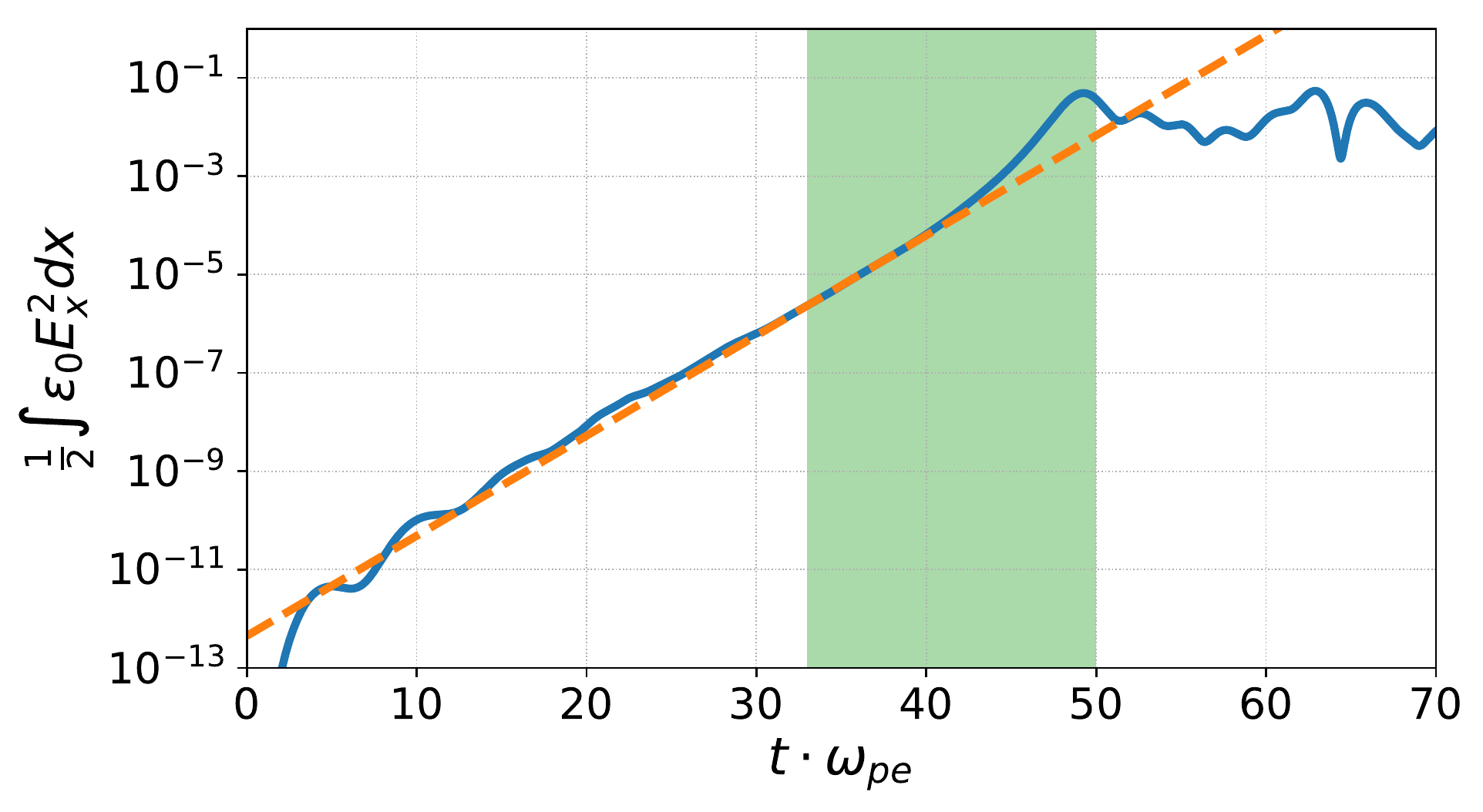}
  \caption{Growth of the electric field energy together with the best
    exponential fit.  The electric field energy growth rate obtained
    is twice the magnetic field energy growth rate, which is
    consistent with Eq. \ref{eq:egrowth}.  Note the growth before
    saturation which is faster than the exponential.}
  \label{fig:egrowth}
\end{figure}

\begin{figure}[!htb]
  \centering
  \includegraphics[width=\linewidth]{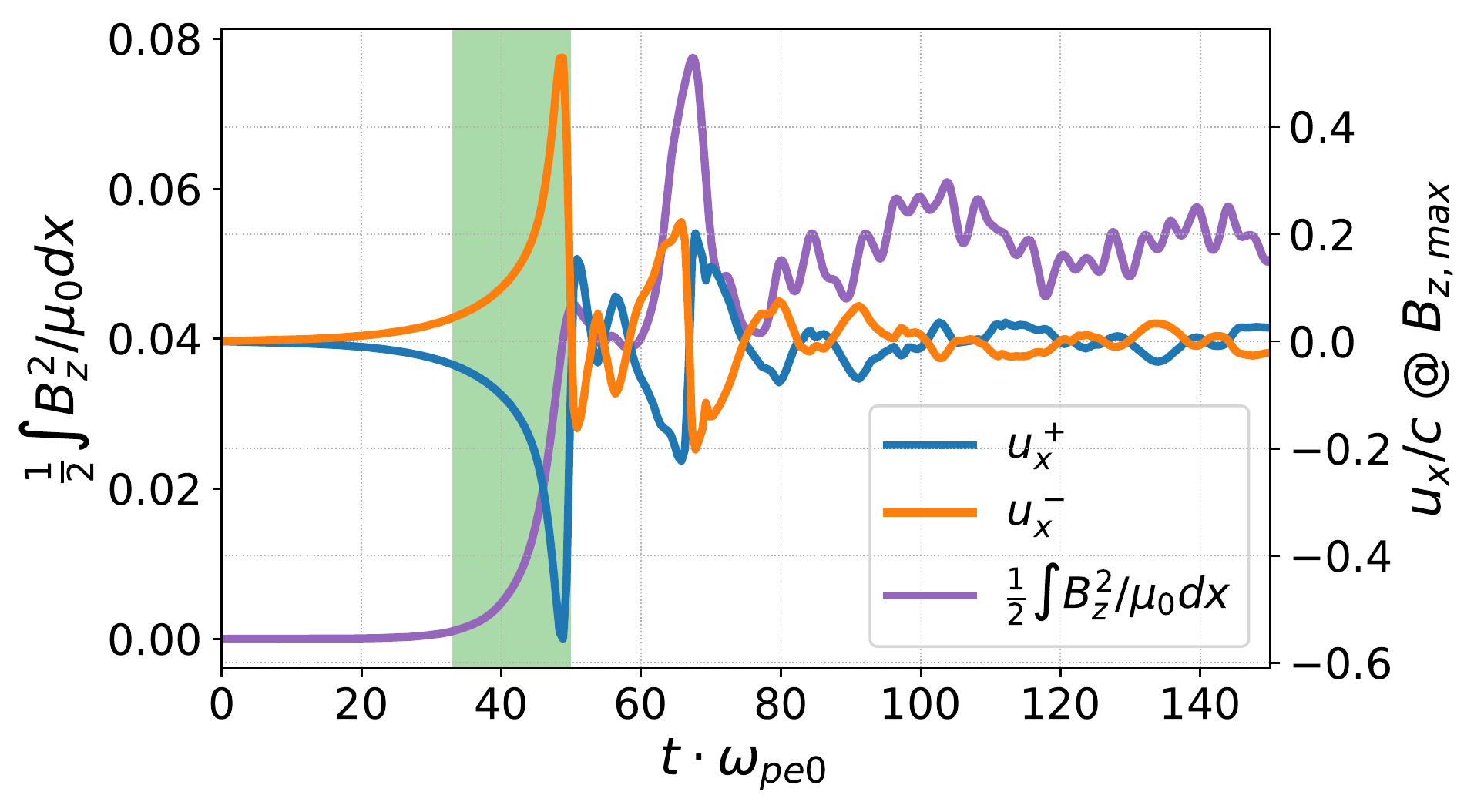}
  \caption{Maximal bulk velocities for both populations together with
    the magnetic field energy growth.  Highlighted in green is the
    area where the two-stream instability dispersion relation
    (Eq.\thinspace\ref{eq:ts}) has a growing root.}
  \label{fig:tsgrowth}
\end{figure}

Fig.\thinspace\ref{fig:egrowth} plots the growth of the electric field
energy as a function of time for the low-temperature case.  Note that
right before saturation (before $t\omega_{pe}\sim 50$), there is an
enhancement in the growth of the electric field energy.
The enhancement of electric field energy growth right before
saturation could be due to a secondary instability.  In this work a
two-stream-like instability is explored.  The green shaded region in
Fig.\thinspace\ref{fig:egrowth} represents the regime where the
transverse velocities ($u_x$), which vary with time, are in a regime
that is unstable to the electrostatic two-stream instability.  The
transverse velocities change rapidly in the $x$-direction hence, there
is not a single classical two-stream growth rate that is relevant to
this regime.  The growth rate of the secondary instability changes
rapidly with time. The dispersion relation of the classical two-stream
instability is given as
\begin{equation}\label{eq:ts}
  1 - \frac{1}{4k^2\lambda_D^2}\left[
    Z'\left(\zeta_1\right) +
    Z'\left(\zeta_2\right)\right] = 0,
\end{equation}
where
\begin{equation}
\zeta_{1,2} = \frac{\frac{\omega}{\omega_0}}{\sqrt{2}k\lambda_D} \pm
    \frac{u_d}{\sqrt{2}v_{th}}
\end{equation}
and $Z'$ is the first derivative of the plasma dispersion function.
The time period over which Equation (\ref{eq:ts}) shows a growing
two-stream instability is denoted in green in
Figs.\thinspace\ref{fig:egrowth} and \ref{fig:tsgrowth}.

\subsection{Phase-space and temperature evolution}

\begin{figure}[!htb]
  \centering
  \includegraphics[width=\linewidth]{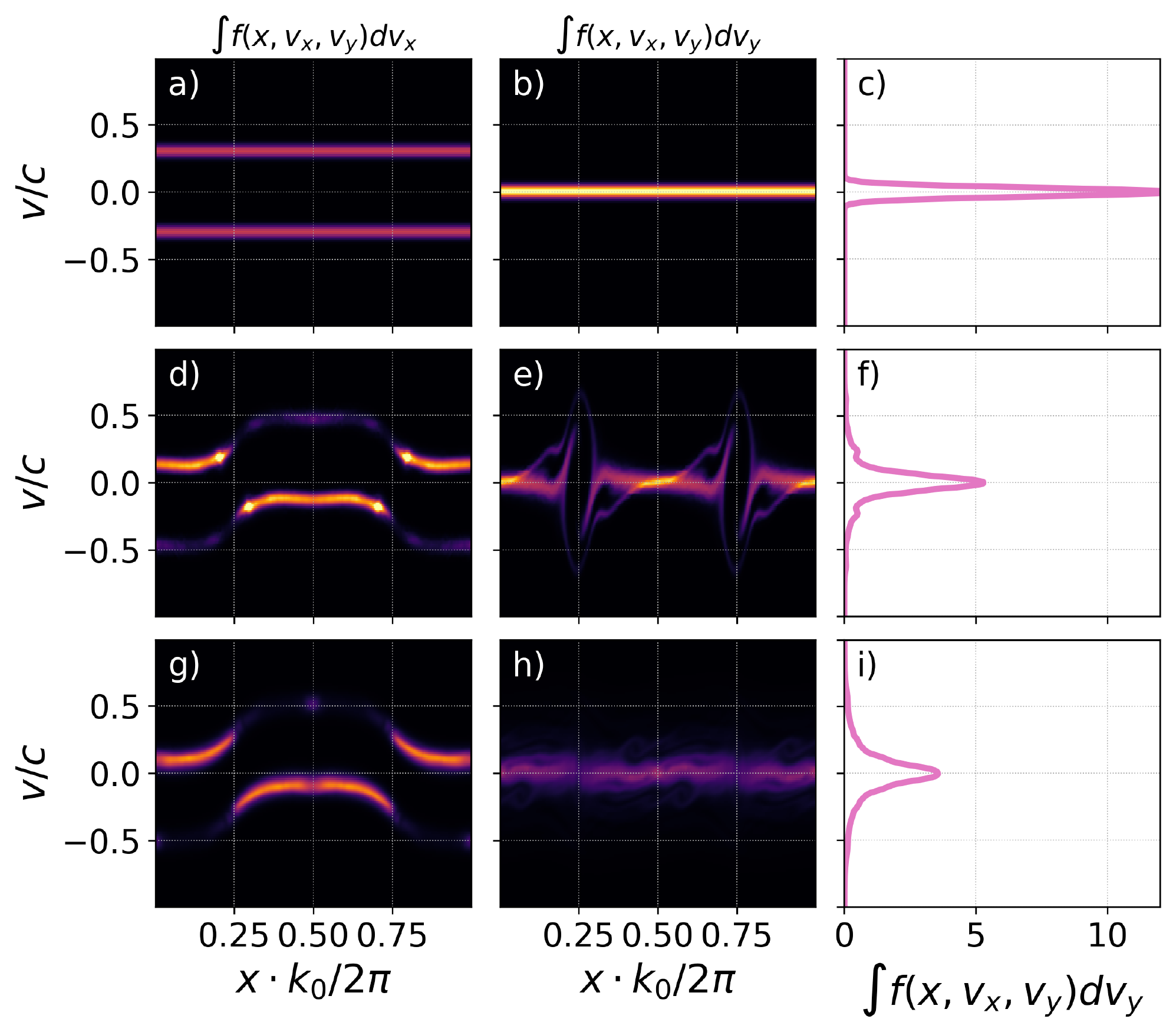}
  \caption{Phase-space plot of the full distribution function
    initially (first row), at the time of first kinetic saturation
    ($t\cdot\omega_{pe0}=50$), and at the end of the simulation
    ($t\cdot\omega_{pe0}=150$). The first column shows distribution
    function integrated with respected to $v_x$, which provides
    insight into $v_y$ structure. Second column shows $x\times v_x$
    distribution (integrated over $v_y$) and captures particle
    trapping vortexes during saturation. Last column contains
    distribution function cross-section integrated in the region
    between the magnetic extremes (from $\frac{1}{4}$ to $\frac{3}{4}$
    of the domain). It shows the overall heating of the electron
    population due to nonlinear phase-mixing.  The results are from
    the simulation with the temperature in between the two extreme
    cases.}
  \label{fig:phasespace}
\end{figure}

The trapping of particles in the potential wells near the magnetic
field peaks is also seen directly in the phase-space plots of the
distribution function in Fig.\thinspace\ref{fig:phasespace} for the
low-temperature case.  In order to present 2D descriptions of the 3D
(1X2V) distribution function, $f(x,v_x,v_y)$ is integrated in $v_x$ to
give $\hat{f}(x,v_y)$ and in $v_y$ to give $\hat{f}(x,v_x)$. The first
row of Fig.\thinspace\ref{fig:phasespace} shows the initial
conditions, the second-row plots these quantities at the time of
kinetic saturation ($t\cdot\omega_{pe0}=50$), and the third row is at
the end of the simulation ($t\cdot\omega_{pe0}=150$). Particle
trapping and phase space mixing are clearly seen in panel (e). The
bright spots are separatrices between the trapped/passing regions. The
last column shows the 1D $v_x$ profiles of the distribution function
integrated over all $v_y$ and averaged over $x$ between the magnetic
peaks (i.e., from $1/4$ to $3/4$ of the domain). Late in time, the
distribution function has significantly broadened due to phase space
mixing, consistent with the earlier description plasma heating during
instability evolution.  The broadening of the distribution function is
also seen in the high-temperature case (not shown here).

Changing the wavelength of the initial perturbation leads (for
unstable wave numbers) to the same qualitative behavior, i.e growth
and saturation of electric and magnetic fields and quasi-periodic
nonlinear behavior due to transverse flow polarity reversal. However,
higher wavenumbers display a much shorter period. This indicates that
the saturation mechanisms observed here are likely universal for the
WI particularly in regimes where the plasma is relatively cold
compared to the drift velocity.  Furthermore, the inclusion of
temperature anisotropy in the counter-streaming populations also
provides consistent results for nonlinear saturation and late-time
nonlinear behavior.

\section{Summary}

The high-order continuum kinetic methods used in this work allow for
noise-free interpretation of detailed plasma dynamics in the kinetic
regime.  Due to their high dimensionality and significant
computational expense, these methods were challenging until recently.
Here, a detailed description of plasma dynamics is presented leading
to the nonlinear saturation of the WI with distribution functions
described well into the nonlinear phase of the instability.  In
agreement with previous work, the results presented here show the
significance of particle trapping due to the magnetic fields.  The
simulation results using $v_{th} = u_d$ confirm magnetic trapping as
the sole mechanism of the instability saturation.  However, this work
additionally emphasizes the role of the electrostatic potential in
regimes where $v_{th} < u_d$.  In the case of cold counter-streaming
plasma beams, the electric field creates potential wells comparable to
the magnetic field potential which significantly modifies the overall
particle trapping.

\begin{acknowledgments}
  Authors would like to thank James Juno for many fruitful
  discussions.  Simulations were performed at the Advanced Research
  Computing center at Virginia Tech (http://www.arc.vt.edu). This
  research was supported by the Air Force Office of Scientific
  Research under grant number FA9550-15-1-0193. The work of Ammar
  Hakim was supported by the U.S. Department of Energy under Contract
  No. DE-AC02-09CH11466.
\end{acknowledgments}




\bibliography{reference}

\end{document}